\documentclass[manuscript]{acmart}

\usepackage[colorinlistoftodos,prependcaption,textsize=small]{todonotes}
\usepackage{subfig}
\usepackage{booktabs}
\usepackage{multirow}
\usepackage{caption}
\usepackage{wrapfig}
\usepackage{enumitem}
\usepackage{tabularx}
\usepackage{array} 

\newcolumntype{C}{>{\centering\arraybackslash}X}

\AtBeginDocument{%
  }

\setcopyright{acmlicensed}
\copyrightyear{2026}
\acmYear{2026}
\acmDOI{XXXXXXX.XXXXXXX}
\acmConference[PEARC '26]{Practice and Experience in
Advanced Research Computing}{July 26--30,
  2026}{Minneapolis, MN}
\acmISBN{978-1-4503-XXXX-X/2026/06}

\begin{document}


\title{Adviser: An Intuitive Multi-Cloud Platform for Scientific and ML Workflows}


\author{Shihan Cheng}
\email{shihan.cheng@vanderbilt.edu}
\orcid{0009-0001-5857-6649}
\author{Michael A.\ Laurenzano}
\email{michael.a.laurenzano@vanderbilt.edu}
\orcid{0009-0001-0370-6725}
\affiliation{%
  \institution{Vanderbilt University}
  \city{Nashville}
  \state{Tennessee}
  \country{USA}
}

\author{Brian Strauch}
\email{brian@adviser.sh}
\author{Timothy A.\ Ellis}
\email{tim@adviser.sh}
\author{Krish Wadhwani}
\email{krish@adviser.sh}
\affiliation{%
 \institution{Adviser Labs, Inc.}
 \city{Atlanta}
 \state{Georgia}
 \country{USA}}

\author{David A.\ B.\ Hyde}
\orcid{0009-0004-4950-5533}
\affiliation{%
  \institution{Vanderbilt University}
  \city{Nashville}
  \state{Tennessee}
  \country{USA}
}
\affiliation{%
  \institution{Adviser Labs, Inc.}
  \city{Atlanta}
  \state{Georgia}
  \country{USA}
}
\email{david.hyde.1@vanderbilt.edu}


\begin{abstract}

Effectively leveraging the vast computational resources of modern cloud environments requires expertise spanning multiple technical domains: configuring scientific software with correct parameters and dependencies, navigating thousands of provider-specific instance types and pricing options, and managing parallel or distributed execution.
We conduct a study indicating that the absence of these categories of expertise poses an ongoing challenge to unlocking the potential of cloud-enabled computational science. To address this challenge, we introduce Adviser, an intuitive multi-cloud platform centered on a workflow abstraction. 
Workflows are reusable, expert-crafted artifacts encapsulating environment setup, data processing, simulation, result capture, and visualization steps needed to execute scientific and ML applications. 
This approach allows users to specify high-level intent, while Adviser handles resource provisioning, runtime configuration, and data movement. 
Using two computational glaciology codes, Icepack and PISM, we show how to use Adviser to gain scientific insight and perform rapid exploration of cost-performance tradeoffs and scaling behavior without specialized expertise in cloud or high-performance computing.

\end{abstract}



\received{20 February 2026}
\received[revised]{30 March 2026}
\received[accepted]{5 June 2026}

\maketitle

\section{Introduction}
\label{sec:introduction}

Cutting-edge computational science workflows increasingly combine large observational datasets with blends of compute-intensive machine learning and conventional simulation techniques. 
Meanwhile, modern cloud environments offer on-demand access to a rich and expanding set of capabilities: CPU, GPU and acceleration hardware; high-throughput networking; tiered object and block storage; managed schedulers; reproducible container-based software primitives; and pay-as-you-go elasticity supporting everything from interactive notebooks to highly parallelized multi-node runs. 

\setlength{\columnsep}{16pt}

\begin{wrapfigure}[10]{r}{0.42\textwidth}
\centering
\includegraphics[width=0.42\textwidth]{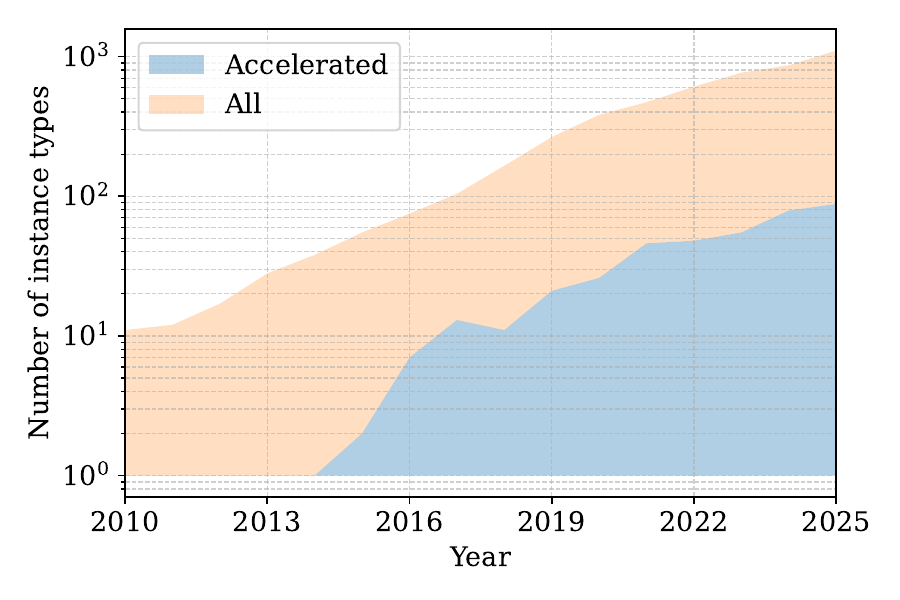}
\vspace{-7mm}
\caption{Exponential growth in launchable Amazon Web Services EC2 instance types over the last 15 years.}
\label{fig:ec2-instance-types}
\end{wrapfigure}

For example, Figure~\ref{fig:ec2-instance-types} shows the number of launchable instance types among both conventional and accelerated compute options within Amazon Web Services (AWS) EC2. 
The number of available options has exploded over the past decade, from dozens to now over 1,000 available choices. 
In principle, this expansive set of choices should result in broadened access to advanced scientific workflows with improved code-to-hardware fit, reduced time-to-solution, lowered cost of experimentation, shortened path to scientific insight, and broader dissemination of results. 
In practice, however,
significant barriers remain to achieving reliable, cost-effective execution at scale to make use of these resources:

\begin{enumerate}[leftmargin=*,itemindent=0pt,wide, labelwidth=!, labelindent=0pt]

\item\textbf{Scientific \& ML Domain Expertise:} running a simulation or model correctly requires domain-specific knowledge rarely captured in one place: e.g., how to load and preprocess datasets, how to install dependencies, and what parameters and settings to use to generate meaningful outputs.
This knowledge is often scattered throughout many papers and scripts, and small mistakes can be difficult to catch while costing users significant overhead.

\item\textbf{Cloud Technology Fluency:} efficient cloud execution requires fluency with hundreds of provider-specific offerings that include instance and accelerator families, networking and storage options, quotas, and pricing options that differ across providers and evolve continuously.
Without sufficient fluency, users often misconfigure, underprovision, or overprovision cloud resources, each of which results in wasted cost or time.

\item\textbf{Distributed Systems Knowledge:} reliable multi-node execution requires HPC and distributed systems expertise that includes MPI/runtime configuration, threading and synchronization, parallel I/O strategies, scaling behavior, fault handling, and reproducibility practices that many users lack.
Compounding the challenges of dealing with distributed systems is that problems often appear only at scale, where debugging is most difficult and inefficiencies are most costly.

\item\textbf{Institutional Access \& Support:} for many potential users, the main barrier to efficient cloud computational science is the lack of access to expertise and support, such as staff familiar with evolving cloud offerings. Users may further lack organizational infrastructure such as cloud credits, software licenses, and training pathways, particularly in settings such as non-R1 universities.

\end{enumerate}

Prior work addresses these barriers through three approaches: science gateways that package domain-specific tools behind curated interfaces~\cite{madhavan2013nanohub, tran2022simvascular, eynard2019pangeo}, cloud orchestration layers that automate provisioning and deployment~\cite{yang2023skypilot}, and workflow systems that help developers express parallelism and compose pipelines~\cite{bauer2012legion, di2017nextflow, deelman2015pegasus}. Each approach addresses only a subset of the barriers: gateways are difficult to extend beyond their target domains, orchestration layers leave users to supply domain and systems expertise, and workflow abstractions require significant implementation effort without providing domain-ready configurations. No existing system systematically encodes all three forms of expertise into reusable, portable artifacts, a gap we target in this work.

This paper introduces Adviser, an intuitive multi-cloud platform carefully designed to address the main barriers to cloud adoption through a unified workflow abstraction. The central idea behind the workflow abstraction is to represent scientific computation as parameterized recipes that package domain-aware setup (datasets, model configuration, preprocessing steps, environment setup), execute portably across providers via established open-source abstraction layers, and standardize the distributed-systems practices required for robust operation (environment capture, structured logging, provenance, and repeatable outputs). Users interact with Adviser through either a CLI or web interface. By encoding domain knowledge, cloud infrastructure expertise, and distributed‑systems best practices into reusable components, Adviser simultaneously addresses technical barriers for individual users and reduces the institutional support burdens that give rise to access barriers. In addition to the Adviser system, which is the core contribution of the paper, this paper contributes:

\begin{itemize}[noitemsep,topsep=0pt]

\item{\emph{An empirical analysis of real‑world job postings} (Section \ref{sec:job-analysis}) characterizing the role of scientific/ML domain expertise, cloud fluency, and distributed-systems knowledge in HPC-related positions.}

\item{\emph{The system architecture of Adviser} (Section \ref{sec:design}) grounded in empirical analysis on observed gaps in domain, cloud and distributed systems expertise and based on reusable templates, environments and execution semantics.}

\item{\emph{End‑to‑end validation in computational glaciology} (Section \ref{sec:usecases}), showing how users can run Icepack and PISM workflows, including parameter studies, MPI‑parallel executions, and visualization, without specialized expertise in glaciology, cloud infrastructure, or distributed systems.}

\end{itemize}

\section{Related Work}
\label{sec:related_work}

\noindent\textbf{Science Gateways \& Domain Platforms.} Science gateways and domain platforms are designed to lower the barriers to advanced computation for scientists and students, particularly in specific science domains. One of the most mature examples is nanoHUB.org, which provides a comprehensive cloud-hosted platform for nanoscale modeling and education, with hundreds of simulation tools, course materials, and detailed usage analytics on classroom and research impact~\cite{madhavan2013nanohub}. 
More recently, the SimVascular Gateway extends this model to cardiovascular simulation: Tran et al.\ describe how a specialized web-based interface, backed by HPC resources, enables both research and training workflows around patient-specific hemodynamics~\cite{tran2022simvascular}. In the geosciences, community efforts such as the Pangeo ecosystem have popularized cloud-native, open-source stacks for large-scale analysis of atmospheric and ocean datasets, emphasizing scalable notebooks, dask-based parallelism, and reproducible, containerized environments~\cite{eynard2019pangeo}. 
In the life sciences, CyVerse makes computational research more accessible by integrating data management, hosted tools, and workflow execution services~\cite{swetnam2024cyverse}. 
Tapis approaches the gateway problem through a reusable, service-oriented API layer, providing hosted interfaces for job execution and data management across distributed resources, along with fine-grained permissions and provenance endpoints that support reproducible computational research~\cite{stubbs2021tapis}.

A complementary thread of domain platforms focuses less on general gateway interfaces and more on packaging complete, ready-to-run scientific environments for specific models and communities. 
For example, prior work has evaluated and operationalized cloud deployments for major Earth and environmental modeling codes, including CESM on AWS~\cite{chen2017running} and a series of GEOS-Chem studies that emphasize immediate access to preconfigured software and data, as well as scalable MPI performance on commercial clouds~\cite{zhuang2019enabling,zhuang2020enabling}. Related efforts demonstrate reproducible, infrastructure-as-code workflows for key codes within important domains such as CFD~\cite{oreproducible} and seismic simulation~\cite{breuer2019petaflop}. 

These systems demonstrate that domain-oriented platforms can broaden access by packaging together the tools, environments, and curated interfaces helpful within each domain. 
Adviser takes a complementary approach, targeting reuse of workflow artifacts, portable environment descriptions, and execution semantics rather than the stack necessary for a particular scientific domain. 
Instead of building a bespoke portal around one community’s models, Adviser represents applications as parameterized workflow templates that bundle domain-ready defaults (data staging, preprocessing, validated configurations) with portable environment specifications, then executes them through a uniform orchestration layer with consistent run semantics and provenance tracking across cloud and  HPC backends. 
This template-centric approach supports both research and instructional settings by facilitating the sharing of expert-vetted workflows and environment setup, controlling permissions and budgets, and reproducing runs.

\noindent\textbf{ML Infrastructure Platforms.} Several commercial platforms in the machine-learning ecosystem reduce infrastructure friction by providing hosted development environments and managed execution for common ML workloads. 
Saturn Cloud offers managed data science workspaces with hosted IDEs, elastic compute, and team collaboration features, enabling rapid onboarding without requiring users to operate their own notebook and cluster infrastructure~\cite{saturn-cloud}. 
NVIDIA run:ai targets efficient use of GPU resources for AI workloads on Kubernetes by providing scheduling and resource sharing mechanisms that improve throughput on expensive accelerators~\cite{nvidia-run-ai}. 
Modal provides a Python-native serverless platform for compute- and data-intensive jobs, allowing users to run batch workloads, training, and inference without maintaining persistent clusters~\cite{modal-stack}. Adviser is complementary to these systems, organizing around end-to-end scientific and ML workflows  rather than focusing solely on ML training and inference tasks. Adviser packages workflows as reusable templates that bundle domain baselines, including datasets, preprocessing, and validated configurations, and produces standardized outputs with recorded provenance supporting reproducibility and comparison across cloud and HPC backends. 
Adviser also emphasizes organizational constraints common in scientific computing, including shared templates, budgets, permissions, and supported configurations that enable teams and classrooms to adopt large-scale computing in a controlled and repeatable manner.

\noindent\textbf{Programming \& Workflow Abstractions.} Complementary work reduces complexity through new programming abstractions and workflow orchestration. PGAS models such as UPC/UPC++~\cite{carlson1999introduction, upcxx_ipdps19} and Co-array Fortran~\cite{numrich1998co} enable fine-grained coordination in parallel code, while task-based systems like Legion~\cite{bauer2012legion}, OmpSs-2@Cluster~\cite{ompss2cluster_europar22}, and Parsl~\cite{babuji2019parsl} express parallelism through logical regions or task graphs that runtimes map to heterogeneous hardware. 
At lower levels, Argobots~\cite{argobots_tpds17} provides lightweight threading primitives, nOS-V~\cite{nosv_ipdps24} explores system-wide scheduling, and IRIS~\cite{irisreimagined_wamta24} coordinates execution across accelerator backends. 
Scientific workflow managers such as Nextflow~\cite{di2017nextflow}, Snakemake~\cite{koster2012snakemake}, and Pegasus~\cite{deelman2015pegasus} offer declarative pipeline specifications with support for containerized execution and cloud backends; cross-language data interfaces like Apache Arrow~\cite{lentner2019shared} reduce serialization overhead across these systems. Adviser does not introduce new programming models or workflow languages; instead, it packages existing applications into reusable templates with portable environments and recorded provenance, allowing users to benefit from advances in programmability and workflow orchestration without programming against them directly.

\section{Barriers to HPC in the Cloud}
\label{sec:job-analysis}

In Section~\ref{sec:introduction}, we hypothesize four barriers that hinder adoption of cloud-based HPC for scientific workflows. Three of these, Scientific \& ML Domain Expertise, Cloud Technology Fluency, and Distributed Systems Knowledge, are technical in nature and can be assessed through explicit signals, communications or products found within the HPC community. On the other hand, institutional barriers by definition exclude individuals and organizations, impeding data collection and empirical analysis; we thus focus on the first three barriers.

To validate that these technical barriers are practically binding, constraining real-world adoption and requiring dedicated expertise to overcome, we leverage HPC industry job postings from the well-known industry forum HPCWire\footnote{\url{https://jobs.hpcwire.com}} as an external signal.
Job postings represent revealed preferences at the organizational level.
Employers recruiting and showing a willingness to pay for particular technical capabilities indicate that those capabilities are necessary for continued growth/success of the organization and that those capabilities are sufficiently scarce that they cannot be assumed as background knowledge.
Insofar as the hypothesized technical barriers appear consistently as required or central qualifications across HPC job postings, this can be taken as evidence that these skills must be explicitly acquired and that the lack of those skills is an impediment to non-experts attempting to run scientific workloads at scale.


\begin{table*}[t]
  \centering
  \small
  \setlength{\tabcolsep}{6pt}
  \renewcommand{\arraystretch}{1.25}
  \begin{tabular}{p{0.06\linewidth} | p{0.42\linewidth} | p{0.44\linewidth}}
  \toprule
  \textbf{Score} & \textbf{Pass 1: Job Technical Relevance} & \textbf{Pass 2: Technical Barrier Criticality} \\
  \midrule
  1 & Definitely not technically relevant & Not mentioned in the job posting \\
  2 & Unlikely to be technically relevant & Could be helpful for performing the role \\
  3 & Possibly technically relevant & Definitely helpful for performing the role \\
  4 & Likely technically relevant & Required for the role \\
  5 & Definitely technically relevant & Central to the role \\
  \bottomrule
  \end{tabular}

  \begin{tabular}{p{0.10\linewidth} | p{0.84\linewidth}}
  \toprule
  \textbf{Pass} & \textbf{Prompt Template} \\
  \midrule
  Technical Relevance & ``You are analyzing job postings to assess technical relevance. Given the job title, employer, and full description, rate how likely the role is to involve hands-on
  work with: (a) writing or modifying code, (b) domain-specific scientific or engineering applications, (c) machine learning workflows, or (d) cloud infrastructure or HPC systems." + <RUBRIC> + <JOB POSTING> + <OUTPUT FORMAT>.\\
  \midrule
  Barrier Criticality & ``You are analyzing job postings to score how essential four skillsets are to the role." + <BARRIER DESCRIPTIONS> + <RUBRIC> + <JOB POSTING> + <OUTPUT FORMAT>\\
  \bottomrule
  \end{tabular}
  \caption{Likert scoring rubrics and prompts for the 2-pass LLM-based textual analysis of HPC job postings. Pass 1 filters for technical relevance, while Pass 2 quantifies barrier criticality represented by each posting.}
  \label{tab:job-analysis-scales}
  \end{table*}

We built and deployed a web scraping tool to harvest all job postings from HPCwire, 
on January 29, 2026. Our scraping approach yielded 363 job postings spanning 88 distinct employers, representing a cross-section of organizations actively seeking HPC-related talent including national laboratories, cloud providers, hardware vendors, research institutions, and commercial HPC service providers. 
We employ the two-pass textual analysis summarized in Table~\ref{tab:job-analysis-scales} using \texttt{llama3.3-70b-instruct} to systematically extract relevant signals from the scraped, unstructured job descriptions. 

\noindent\textbf{Pass 1: Technical Relevance Filtering.} The first pass assesses each posting's technical relevance by evaluating how likely the role is to interface with code, domain science, machine learning, or cloud/HPC systems on a 5-point Likert scale. 
We retain postings scoring 3 (``possible/unclear''), 4 (``likely''), or 5 (``definitely''), discarding roles scoring 1 (``definitely not'') or 2 (``unlikely'') that are not expected to involve hands-on technical work with computational infrastructure. 
This filtering pass reduces the corpus from 363 to 201 technically-relevant job postings, removing administrative, sales, and other non-technical positions that are unlikely to be personally making use of HPC applications.

\noindent\textbf{Pass 2: Barrier Criticality.} The second pass evaluates each posting against the three technical barrier categories, assessing on a 5-point scale whether expertise in each category will enable the candidate to successfully perform the role, assigning scores of 1 (``not mentioned in the job posting''), 2 (``could be helpful''), 3 (``definitely helpful''), 4 (``required for'') or 5 (``central to'') to each posting. The purpose of the second pass is to distinguish between skills that are incidental versus those that are core, defining characteristics of the position, allowing us to quantify the criticality of each of the barriers and the expertise required in each job description to overcome them.


\begin{figure}[t]
  \centering
  \includegraphics[width=0.95 \textwidth]{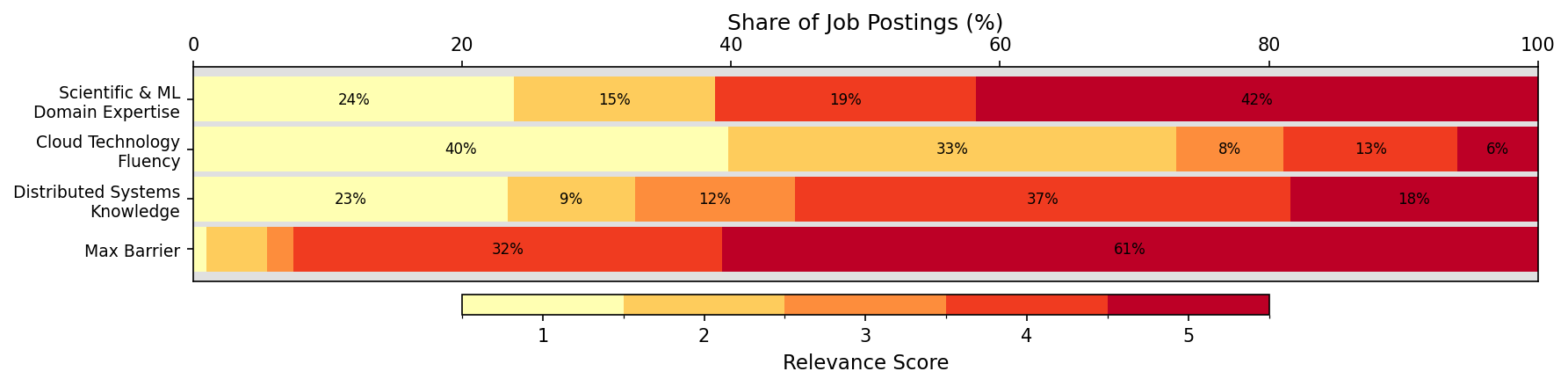}
  \vspace{-4mm}
  \caption{Distribution of barrier criticality scores across 201 technically-relevant HPCWire job postings. Each barrier category (Scientific \& ML Domain Expertise, Cloud
  Technology Fluency, Distributed Systems Knowledge) is scored on a 1--5 scale indicating how critical that expertise is for the role. ``Max Barrier'' shows the distribution of the maximum score across all three barriers for each posting.}
  \label{fig:job-relevance-scores}
\vspace{-2mm}
\end{figure}

Figure~\ref{fig:job-relevance-scores} presents the distribution of barrier criticality scores across all 201 technically-relevant postings for our textual analysis, revealing patterns consistent with the barriers hypothesized in Section~\ref{sec:introduction}. Scientific \& ML Domain Expertise emerges as the most consistently demanded skill, with 61\% of postings rating it as ``required for'' or ``central to'' the role. Distributed Systems Knowledge is also required by a plurality (55\%) of roles. 
Cloud Technology Fluency, while receiving the lowest scores, still appears at the ``definitely helpful'' level or above in 27\% of postings—a figure likely to grow as cloud adoption accelerates. These results confirm that all three barriers represent genuine obstacles requiring dedicated personnel to address, rather than skills that can be treated as incidental or easily acquired by domain scientists and technical generalists.
   
\emph{The vast majority of positions require expertise in at least one barrier area.} The ``Max Barrier'' distribution in Figure~\ref{fig:job-relevance-scores} shows the maximum barrier score across all three categories for each posting. 
The concentration of high scores, where at least one of the three technical barriers is ``required for'' or ``central to'' 93\% of jobs, indicates that the overwhelming majority of technically-relevant HPC positions require dedicated expertise to overcome at least one of the hypothesized barriers. 
This finding strongly supports the premise that these barriers reflect the reality of conditions in the field and represent genuine constraints that organizations are willing to address through dedicated resource investment.

\section{System Architecture}
\label{sec:design}

\begin{figure}[!t]
    \centering
    \includegraphics[width=1.0 \textwidth]{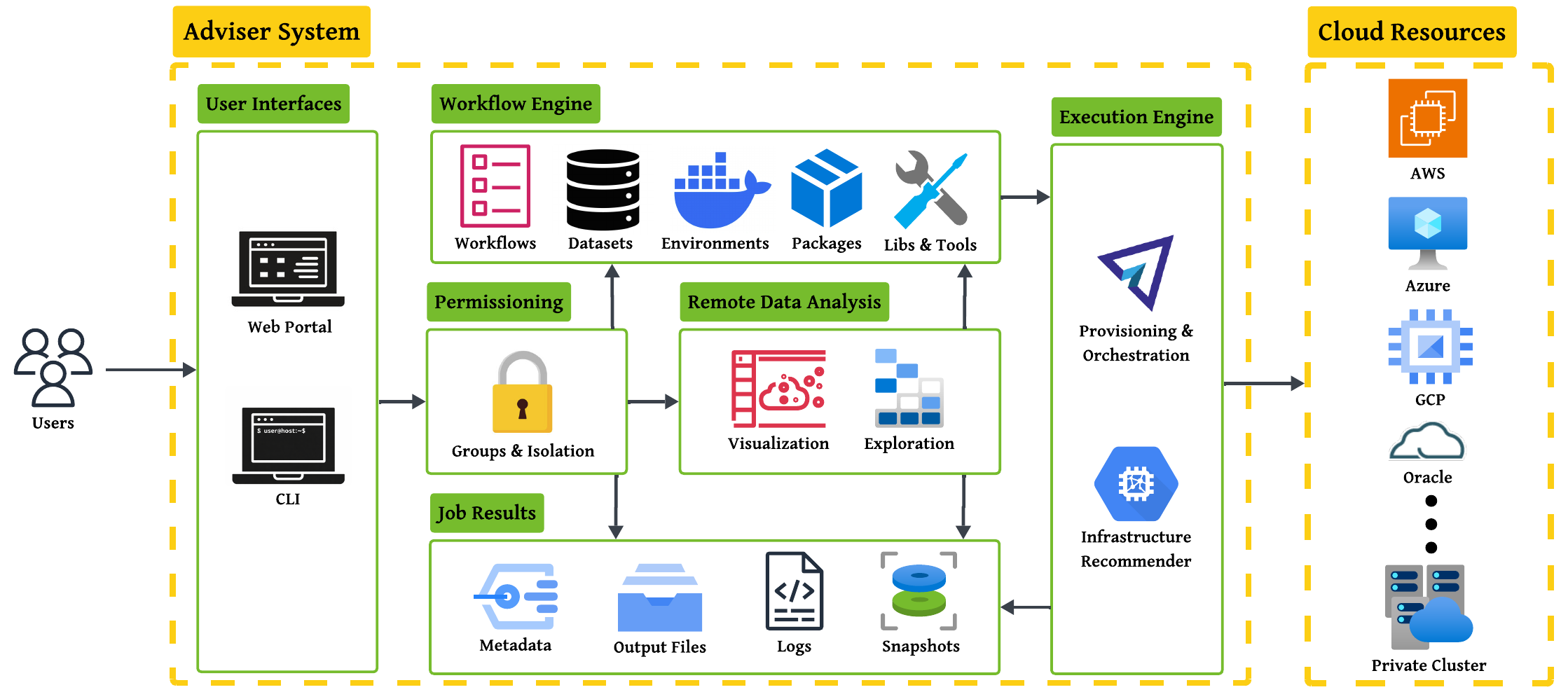}
    \vspace{-6mm}
    \caption{Architectural overview of the Adviser platform highlighting the functionality and interactions of its subsystems.}
    \label{fig:overview}
\vspace{-2mm}
\end{figure}

The analysis in the prior section ultimately informed our design of Adviser, which aims to provide a platform for overcoming the technical barriers outlined in Section \ref{sec:introduction}. 
Adviser is composed of five tightly-integrated subsystems that cooperate to provide intuitive access to reusable scientific workflow artifacts and facilitate seamless, reliable, performant execution of scientific and ML codes on cloud resources for users with expertise in none of the underlying technological domains. 
The structure of Adviser is illustrated in Figure~\ref{fig:overview}. 
Domain expertise barriers are addressed primarily by the Workflow Engine (through curated templates and validated baselines) and reinforced by Job Results (through provenance and diagnostics that make correctness checkable). 
Cloud technology barriers are handled by the Execution Engine (portable provisioning, backend selection, and cost/performance guidance) with supporting abstractions exposed through the User Interfaces. 
Distributed systems barriers are mitigated through standardized execution envelopes cataloged within the Workflow Engine and deployed by the Execution Engine and through consistent observability in Job Results, which reduces the effort required to debug problems that only appear at scale. 
Institutional access barriers are addressed at the boundaries of the system---primarily via User Interfaces and Workflow Engine constructs that support shared workspaces, expert-curated approved configurations, and repeatable onboarding pathways---while also benefiting from Execution Engine mechanisms that enforce budget-aware and policy-conformant execution.

\subsection{User Interfaces}

The User Interfaces within Adviser are designed to lower friction for adoption and use, providing succinct specification of simulation tasks while allowing experts full flexibility when needed. 
As shown in Figure~\ref{fig:overview}, the UI layer consists of a web portal and command-line interface that present a uniform entry point for launching, monitoring, and iterating on workflows. 
Rather than exposing provider-specific configuration knobs, the interfaces emphasize user intent: selecting a workflow template, providing domain parameters, choosing a goal (e.g., quick test, production run, or visualization session), and optionally specifying constraints such as budget or preferred backends. 
This separation reduces cloud technology barriers by avoiding early commitment to provider-specific options, while also addressing institutional access constraints through consistent onboarding workflows (e.g., joining a workspace, discovering available templates/datasets, and running within organization-defined guardrails). 
The UI layer facilitates education and team use: templates and results are surfaced through a uniform interface to significantly lower the activation energy for non-experts, including students and new group members, without requiring them to grapple with the full complexity of the underlying scientific problem, distributed execution, or the underlying cloud resources being leveraged on their behalf.

\noindent\textbf{Capabilities.} The system incorporates a group-based permissioning mechanism that controls access to all resources: workflows, datasets, environments, job results and computational resources. In classroom deployments, instructors can allocate a shared cloud budget, monitor student progress, and distribute standardized workflow templates that provide a consistent learning experience. For industry and research teams, Adviser supports collaborative debugging, shared visualization sessions, and reproducible execution environments that ensure consistency across team members.

\noindent\textbf{Command-Line Interface.} Adviser's CLI exposes the platform's capabilities through a small set of composable commands that are mirrored by capabilities in the web interface. The primary entry point to the end user is \texttt{adviser run}, which accepts a workflow to execute, the workflow configuration, and optional resource specifications:

{
\footnotesize{
\begin{verbatim}
# Set up and run a PISM simulation with automatic instance selection
> adviser run --setup "./setup_pism.sh" "./run_pism.sh"

# Run ML training using specific hardware capabilities without referencing any cloud provider or instance type
> adviser run "python train.py" --gpu 1 --ram 32

# Scaled PISM run for 4-node MPI environment on AWS
> adviser run --setup "./setup_pism.sh" "./run_pism.sh --np 96" --cloud aws --num-nodes 4 --instance-type hpc7a.12xlarge
\end{verbatim}
}
}

The three examples above illustrate how Adviser systematically reduces each technical barrier identified in Section~\ref{sec:introduction}. 
The first command shows how the \texttt{-{}-setup} flag mitigates the domain expertise barrier: the user references a setup script that encapsulates model-specific dataset paths, preprocessing steps, dependency installation, and validated parameter defaults without needing to understand the intricacies of how the software works or the datasets being used. 
The second command shows how Adviser reduces the cloud technology fluency barrier: rather than navigating dozens of GPU-enabled instance types across providers, the user requests capabilities as \texttt{-{}-gpu 1 -{}-ram 32} and Adviser \textit{automatically selects an appropriate and available provider, region and instance type} meeting the infrastructure requirements such as \texttt{g6.2xlarge} on AWS. 
The third command illustrates how explicit infrastructure parameters \texttt{-{}-cloud}, \texttt{-{}-num-nodes} and \texttt{-{}-instance-type} remain available for fine-grained control of resources, while \texttt{-{}-np 96} in the PISM workflow (see Section \ref{sec:usecases}) simplifies the distributed systems interaction: users specify the desired MPI rank count as part of their run command, and Adviser handles hostfile generation, slot allocation, EFA configuration, and runtime environment setup steps that may otherwise require substantial distributed computing expertise to enable.

\subsection{Curated, Reusable Artifacts}

The Workflow Engine subsystem shown in Figure~\ref{fig:overview} is Adviser's primary knowledge center, packaging and cataloging the artifacts needed for repeatable, transferable workflows: scientific codes, run commands, datasets, environment descriptions such as container images or configuration scripts. The core role of the Workflow Engine is to centralize scattered domain know-how into explicit, versioned baselines that can be executed, maintained and audited. 
Workflow templates encode command line directives and domain-relevant configuration including dataset choices, preprocessing steps, parameter defaults, exploration pathways and validation checks that allow common failure modes to be caught early. 
By making these artifacts explicit and shareable, the Workflow Engine directly targets domain expertise barriers and reduces the risk that small configuration mistakes lead to wasted time and cost. 
The engine maintains and uses a curated set of shared libraries, container environments and common tooling providing common functionality across scientific domains, including distributed I/O, checkpointing routines, high performance numerical solvers, and GPU acceleration primitives. 
These libraries deduplicate effort among applications and facilitate performance portability: the same workflow template can run on a single GPU instance, a multi-node CPU cluster, or a hybrid cloud–HPC configuration without modification. 
Environment descriptions establish a portable runtime contract that limits drift across providers and clusters, mitigating cloud technology barriers by decoupling the tools needed by the workflow from the particulars of how an execution environment is assembled and used on a specific set of resources.

\subsection{Cloud-Agnostic Provisioning and Orchestration}

The Execution Engine subsystem shown in Figure~\ref{fig:overview} is the control plane for selecting, provisioning and running workflows across real cloud execution resources. It consumes Workflow Engine artifacts and infrastructure descriptions of available hardware resources then launches distributed jobs and captures execution-time signals needed for monitoring and post-run analysis. 
Adviser uses SkyPilot~\cite{yang2023skypilot} within this subsystem to provide cloud-agnostic provisioning and orchestration, but the design goal is conceptual: users and templates specify requirements at the level of capabilities and constraints (e.g., GPUs, memory, interconnect expectations, budget caps), while the Execution Engine maps those intents to concrete resource selections and backend deployments. 
This mapping is where cloud technology barriers are reduced most directly, as infrastructure complexity and churn are absorbed centrally rather than being dealt with by individual workflow maintainers. 
The subsystem also addresses distributed systems barriers by providing a standardized execution envelope for multi-node runs—consistent runtime configuration, structured logs, and predictable job lifecycle behavior—so that scale-induced failures and inefficiencies are easier to detect and correct.

\subsection{Job Results and Provenance}

The Job Results subsystem in Figure~\ref{fig:overview} is Adviser’s persistent record of computation, storing job metadata/provenance, logs, output files, and application snapshots written by the Execution Engine. These records serve two purposes. 
First, they reduce distributed systems expertise needed by making failures and performance pathologies diagnosable: provenance links all logs and results to a template version, environment specification, parameters, and resource configuration, enabling systematic comparison across runs and backends. 
Second, they ameliorate domain expertise barriers by making correctness more checkable and shareable: teams can compare standardized diagnostics, reproduce baseline runs, and incrementally modify parameters while retaining a clear picture of the specific configuration of runs and changes across runs. Built atop these records and inspired by established server-side visualization approaches~\cite{ahrens2005paraview, rothwell2022quantifying}, Adviser supports remote data visualization and exploration sessions that leverage workflows to perform rendering close to the data on cloud-provisioned resources while offering data and result interactivity through its User Interfaces.

\section{Scientific Workflows Enabled by Adviser}
\label{sec:usecases}

\begin{figure}[t]
    \centering
    \subfloat[Time-to-solution per run]{%
        \includegraphics[width=0.48\textwidth]{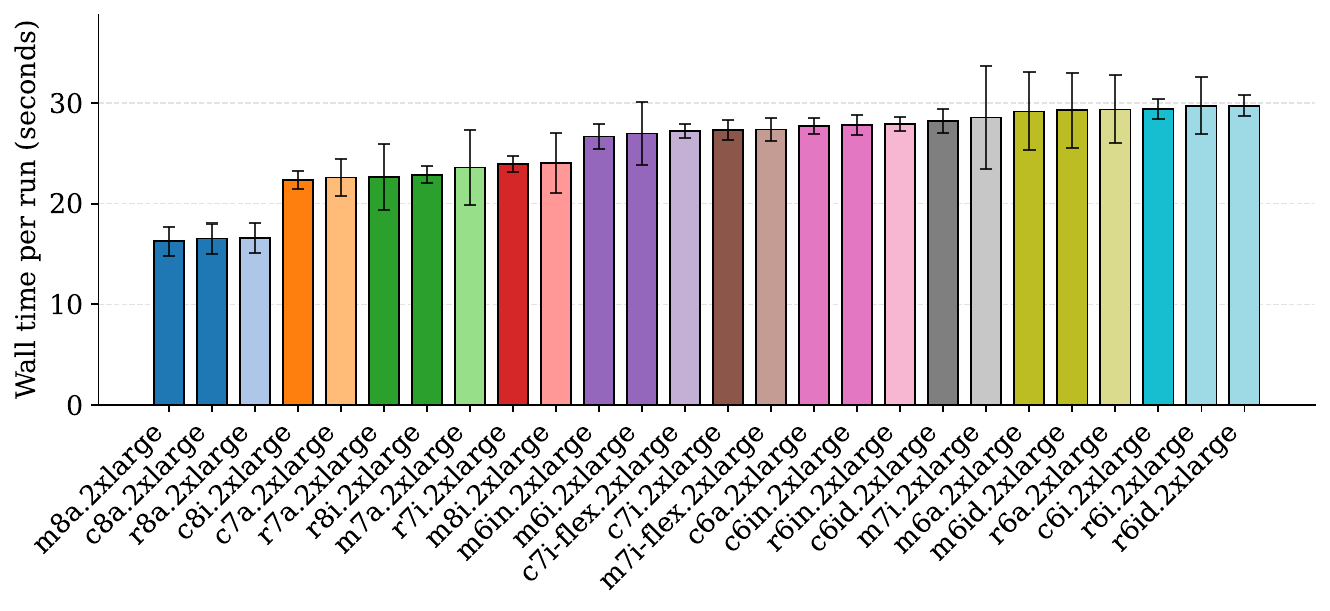}%
        \label{fig:icepack-time}%
    }
    \hfill
    \subfloat[Cost per run]{%
        \includegraphics[width=0.48\textwidth]{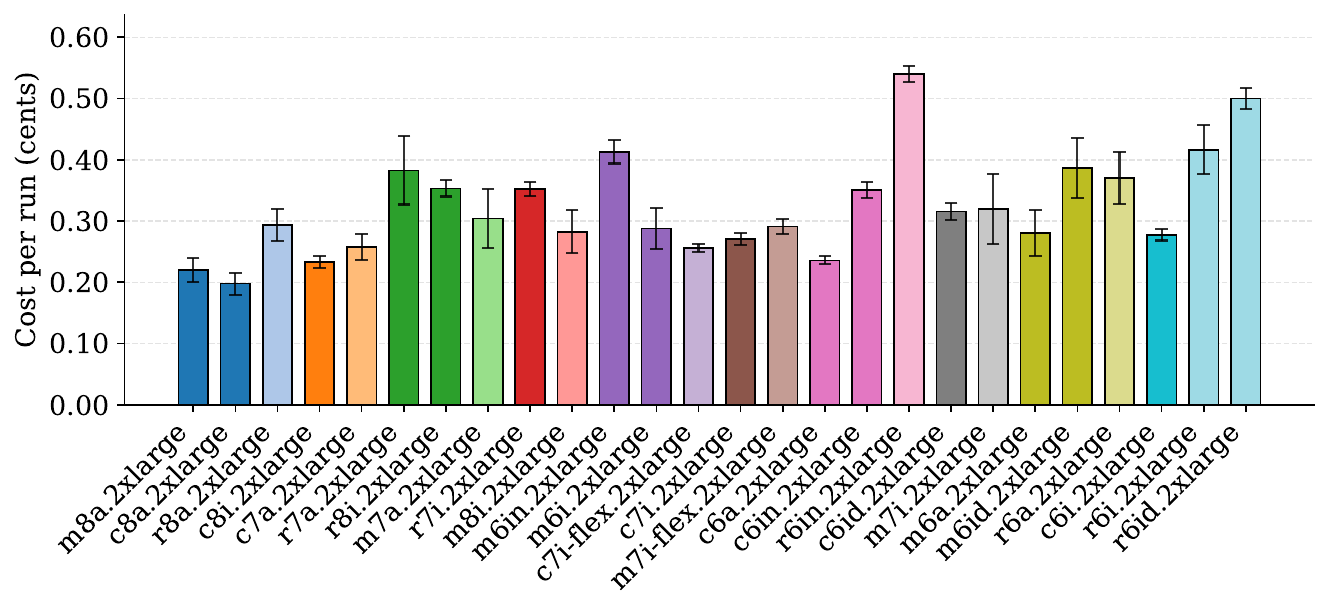}%
        \label{fig:icepack-cost}%
    }

    \vspace{-3mm}
    \caption{Icepack performance and cost characteristics for the synthetic ice shelf workflow across instance types. Bar height represents mean and error bars represent standard deviation over 20 iterated experiments.}
    
    \label{fig:icepack-time-cost}
\vspace{-2mm}
\end{figure}

We illustrate Adviser's ease of use and capabilities through two demonstrations in a relevant domain, computational glaciology, where practitioners often face the barriers discussed in Sections \ref{sec:introduction} and \ref{sec:job-analysis}.


\subsection{Icepack: Firedrake-Based Glacier Dynamics Modeling}

Icepack~\cite{shapero2021icepack} is a Python finite-element framework for glacier modeling built on Firedrake~\cite{rathgeber2016firedrake}. Its workflows require complex environments (MPI, PETSc, platform-specific compilation) that make reproducible deployment nontrivial; Adviser encapsulates these complexities within a reusable workflow to enable single-command launch.

\noindent\textbf{Synthetic Ice Shelf Workflow.} We use Adviser with an idealized 2D ice-shelf flow for a procedurally generated domain with analytically specified thickness and velocity. Each trial uses fixed mesh resolution ($dx=1000$m) and performs a diagnostic solve over $100$ simulated years. We run the same 4-rank MPI configuration across different 8-vCPU (\texttt{2xlarge}) AWS instance types using Adviser's \texttt{-{}-instance-type} option, performing one warm-up step followed by $20$ measured repetitions. We report end-to-end wall-clock time and estimate cost per run using on-demand pricing. Results are summarized in Figure~\ref{fig:icepack-time-cost}, where Figure~\ref{fig:icepack-time-cost}(a) shows that this configuration benefits from successive processor generations: on AMD instances (\texttt{XXa}), time-to-solution decreases from $29.2$s (\texttt{m6a}) to $23.6$s (\texttt{m7a}) to $16.3$s (\texttt{m8a}). Performance is flat across memory tiers within a processor generation (e.g., $16.5$s, $16.3$s and $16.6$s on \texttt{c8a}, \texttt{m8a} and \texttt{r8a}, respectively), suggesting lower-memory configurations reduce cost. Figure~\ref{fig:icepack-time-cost}(b) confirms this intuition: compute-optimized instances (\texttt{cXX}) are cheapest per solution, followed by general-purpose (\texttt{mXX}), then memory-optimized (\texttt{rXX}).

\begin{figure}[t]
    \centering
    \subfloat[Basal melt rate ($\frac{m}{yr}$)]{%
        \includegraphics[width=0.316\textwidth]{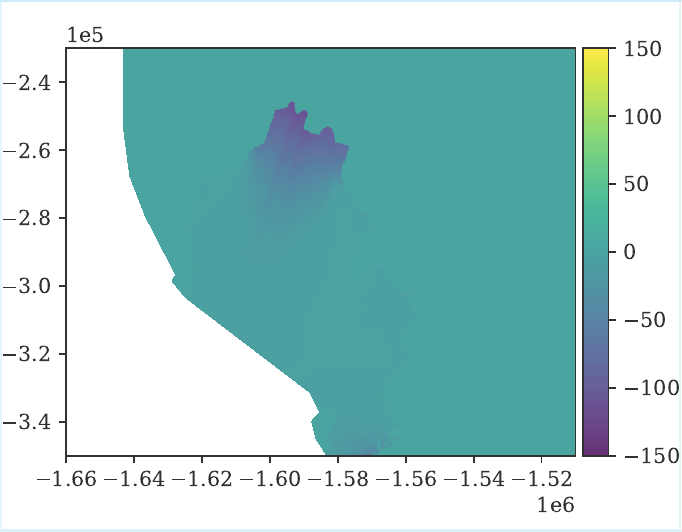}%
        \label{fig:icepack-pine-island-melt}%
    }
    \hfill
    \subfloat[Floating vs. grounded ice ($m$)]{%
        \includegraphics[width=0.304\textwidth]{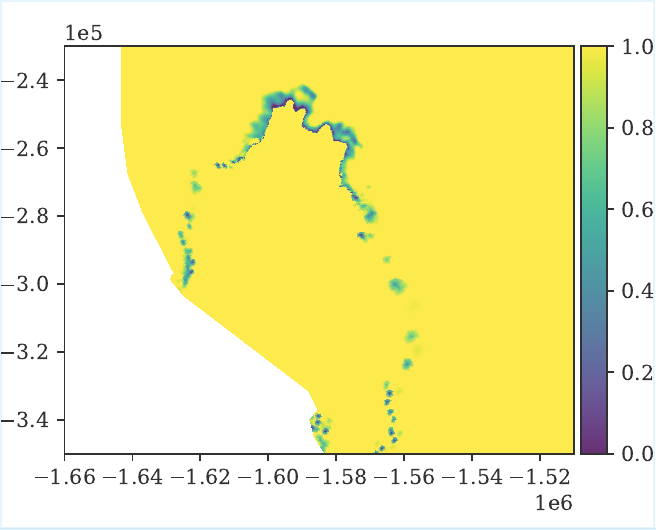}%
        \label{fig:icepack-pine-island-mask}%
    }
    \hfill
    \subfloat[Ice thickness change ($\frac{m}{yr}$)]{%
        \includegraphics[width=0.302\textwidth]{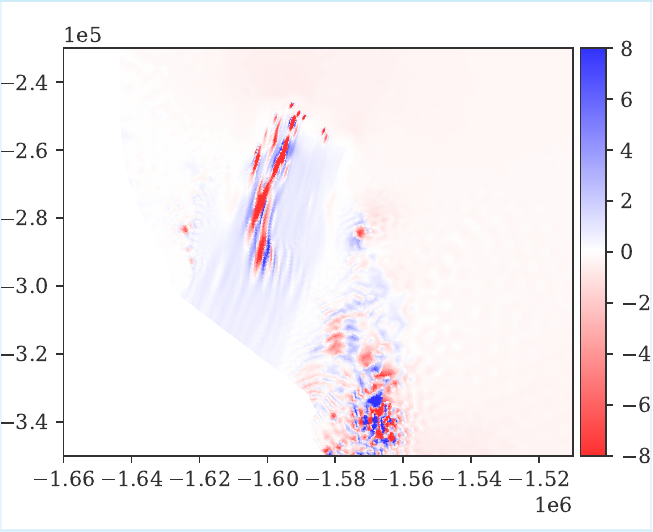}%
        \label{fig:icepack-pine-island-thickness}%
    }
    \vspace{-3mm}
    \caption{Outputs of the Adviser visualization workflow for the Icepack Pine Island Glacier model after 200 simulated years.}
    
    \label{fig:icepack-pine-island}
\vspace{-2mm}
\end{figure}

\noindent\textbf{Pine Island Glacier Workflow.} We use Adviser to simulate the Pine Island Glacier, a major West Antarctic outlet glacier studied for its sea-level rise contribution~\cite{joughin2021ocean, joughin2021ice}. The simulation has two stages encapsulated by Adviser workflows: an inversion step that estimates physical parameters (e.g., basal friction) from velocity observations, producing a calibrated initial state, and a forward simulation that evolves glacier thickness and geometry over 200 years under prescribed surface mass balance and basal melt forcing, solving velocity diagnostically and updating thickness prognostically with floating/grounded regions determined dynamically via flotation criteria. This workflow supports multiple glacier geometries (2017 and 2020 ice-shelf front configurations) and represents a non-trivial scientific workload with complex dependencies, multi-stage execution, and domain-specific outputs. Execution produces structured outputs including spatial fields and visualizations; artifacts are written with post-processing specifications that extract diagnostics automatically. Diagnostic outputs are shown in Figure~\ref{fig:icepack-pine-island}: basal melt rate ($\frac{\text{m}}{\text{yr}}$) in (a), floating versus grounded ice mask indicating grounding line geometry in (b), and ice thickness change rate ($\frac{\text{m}}{\text{yr}}$) in (c). These demonstrate Adviser orchestrating a complex simulation-to-visualization pipeline for a realistic scientific workflow.

\subsection{Parallel Ice Sheet Model (PISM)}

PISM~\cite{winkelmann2011potsdam,martin2011potsdam} is a hybrid shallow-ice/shallow-shelf (SIA+SSA) glaciology model used widely in cryospheric modeling and sea-level applications. Installing PISM reproducibly is notoriously challenging due to its multilayered dependency tree spanning build tools (CMake), scientific libraries (PETSc, NetCDF, FFTW, GSL, UDUNITS), and a Python environment with its own requirements. The stack is also fragile: PISM's documentation warns of failure scenarios when multiple MPI implementations exist, recommending users uninstall all but one MPI library\footnote{\url{https://www.pism.io/docs/installation/prerequisites.html}}, a destructive and impractical step in many environments. The Adviser–PISM artifacts encapsulate setup and execution procedures into a compact workflow, enabling users to run validated PISM configurations without manual environment and installation management.

\begin{figure}[t]
    \centering
    \subfloat[Ice sheet surface elevation $usurf$ ($m$)]{%
        \includegraphics[height=0.30\textwidth]{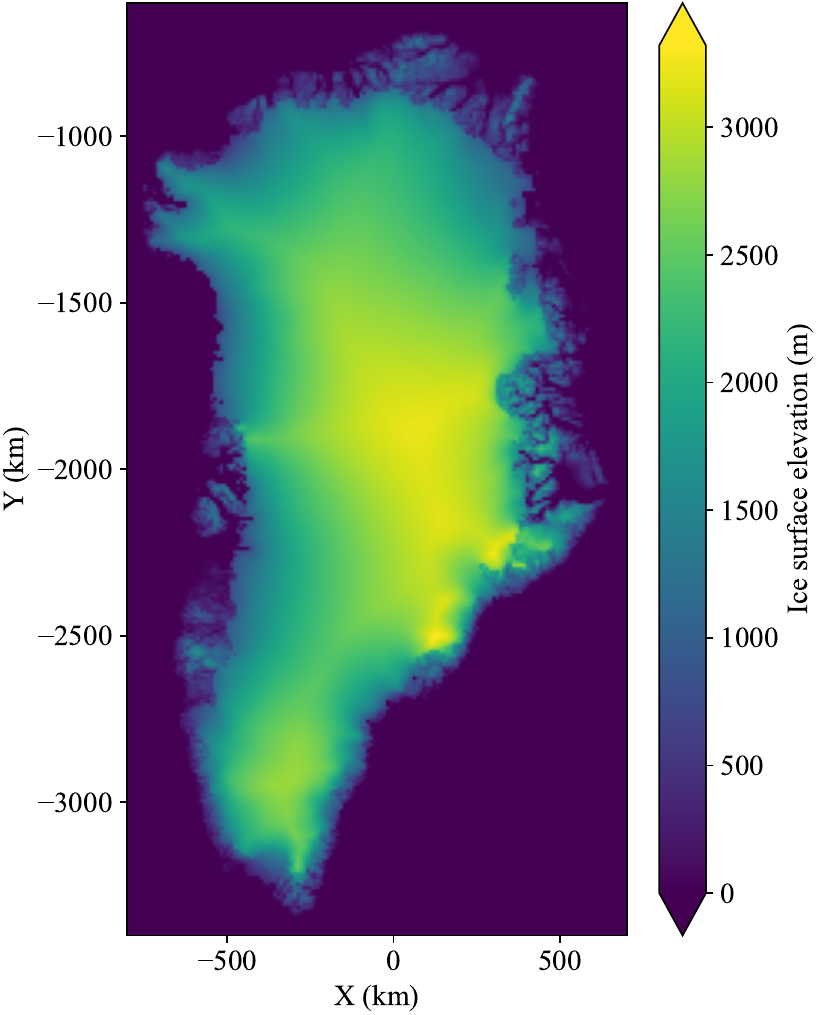}%
        \label{fig:pism-usurf}%
    }
    \hfill
    \subfloat[Surface velocity magnitude $velsurf\_mag$ ($\frac{m}{yr}$)]{%
        \includegraphics[height=0.30\textwidth]{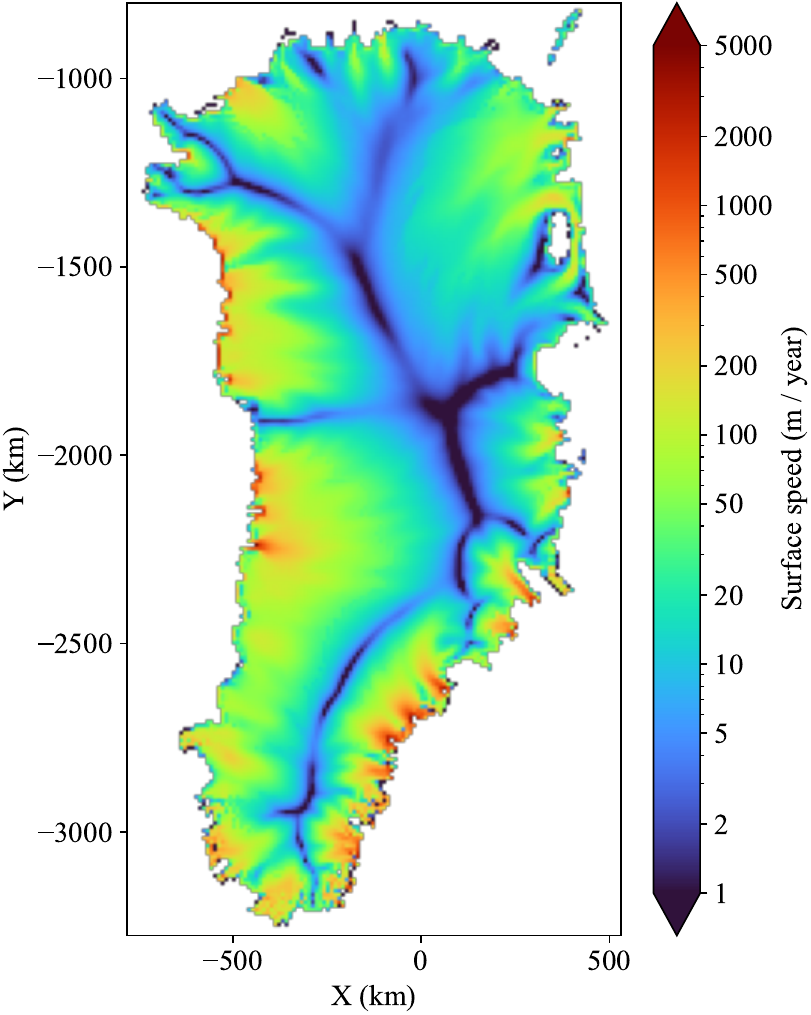}%
        \label{fig:pism-velsurf}%
    }
    \hfill
    \subfloat[Basal sliding velocity $velbase\_mag$ ($\frac{m}{yr}$)]{%
        \includegraphics[height=0.30\textwidth]{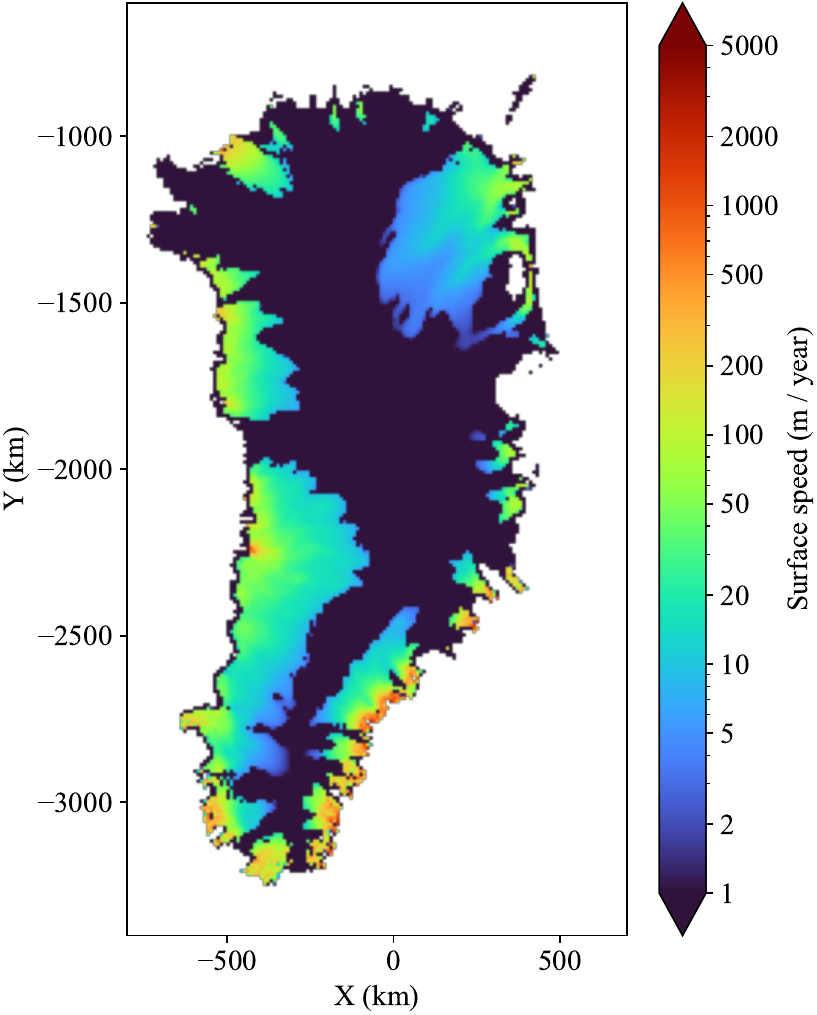}%
        \label{fig:pism-velbase}%
    }
    \hfill
    \subfloat[Land, grounded ice and ocean regions]{%
        \includegraphics[height=0.30\textwidth]{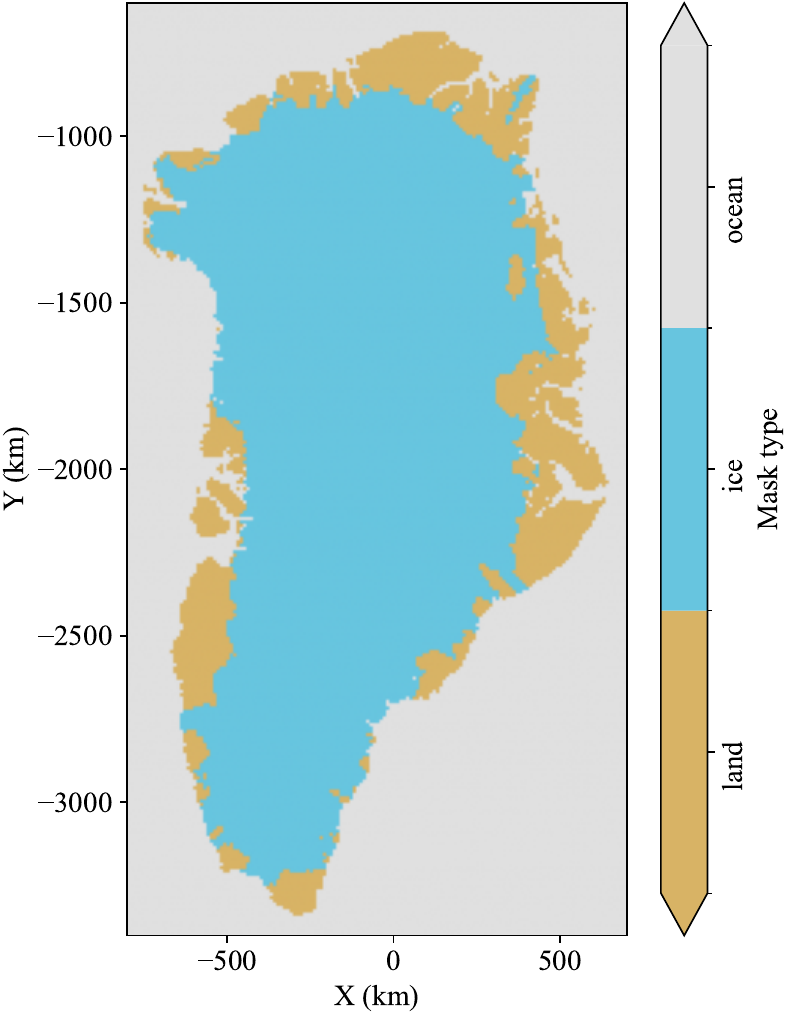}%
        \label{fig:pism-mask}%
    }
    
    \caption{Visualization outputs of Adviser workflow for PISM Greenland simulation.}
    \label{fig:pism-greenland}
\end{figure}

\noindent\textbf{Greenland Workflow \& Visualizations.} We use Adviser to run a Greenland-scale ice-sheet spin-up simulation derived from the PISM manual\footnote{https://www.pism.io/docs/manual/std-greenland/index.html}. The simulation is initialized via PISM's bootstrapping procedure and run under MPI parallelism from $10$kyr BP to present (i.e., year $-10000$ to $0$) using $10$km horizontal grid spacing, constant-climate surface forcing, and PISM's SIA+SSA dynamics model. To demonstrate Adviser's parameter injection capability, we override the default pseudo-plastic sliding law exponent from $q=0.25$ to $q=0.5$ using a single configuration parameter override, simulating more linear sliding behavior and producing a present-day state suitable for subsequent experiments. The resulting diagnostic visualization output of ice sheet surface elevation \textit{usurf}, surface velocity magnitude \textit{velsurf\_mag}, basal sliding velocity \textit{velbase\_mag} and surface-type mask (land/ice/sea) are shown in Figure~\ref{fig:pism-greenland}, validating that Adviser can execute a full ice-sheet spin-up workflow and produce domain-standard model fields at scale.

\noindent\textbf{Scale-up vs. Scale-out Performance.} We measure PISM strong scaling using the Greenland workflow within Adviser, varying MPI ranks from 8 to 96 at fixed 10km resolution with $q=0.25$. All runs use AWS \texttt{hpc7a} instances (4th Gen AMD EPYC, EFA network stack, 300 Gb/s interconnect). We compare two strategies: \textit{Scale-up} uses a single \texttt{48xlarge} (96 vCPUs) with \texttt{-{}-num-nodes 1} at increasing rank counts; \textit{Scale-out} distributes ranks across multiple \texttt{12xlarge} nodes (24 vCPUs each): 1 node for $np \le 24$ (\texttt{-{}-num-nodes 1}), 2 nodes for $np=32,48$ (\texttt{-{}-num-nodes 2}), and 4 nodes for $np=64,96$ (\texttt{-{}-num-nodes 4}). Results in Table~\ref{tab:strong_scaling} reveal distinct scaling profiles. Both achieve ${\sim}1.36$ hours at $np=8$, but Scale-up maintains $33.5\%$ parallel efficiency at $np=64$ with a minimum runtime of $0.5159$ hours. Scale-out efficiency drops sharply beyond 1 node, falling to $13.8\%$ at $np=96$ with no runtime benefit past $np=64$, likely due to inter-node communication latency outweighing the benefit of additional compute, making the single-node architecture the more cost-effective strategy. Adviser enabled seamless switching between these architectures through limited run command modifications, allowing straightforward exploration of resource choices without modifying environment or code.

\begin{table}[htbp]
\centering
\caption{Performance scalability of simulation for the PISM Greenland workflow. Scale-up experiments utilize a single \texttt{hpc7a.48xlarge} instance; scale-out experiments utilize a cluster of \texttt{hpc7a.12xlarge} instances.}
\label{tab:strong_scaling}
\vspace{2mm}
\begin{tabularx}{\textwidth}{l|l|CCCCCCC}
    \toprule
    \textbf{Scaling} & \multirow{2}{*}{\textbf{Metric}} & \multicolumn{7}{c}{\textbf{Number of MPI Ranks ($np$)}} \\
    \cmidrule(lr){3-9}
    \textbf{Strategy} & & \textbf{8} & \textbf{16} & \textbf{24} & \textbf{32} & \textbf{48} & \textbf{64} & \textbf{96} \\
    \midrule
    \multirow{4}{*}{\textbf{Scale-up}} 
    & Time (h)        & 1.38 & 0.80 & 0.87 & 0.71 & 0.56 & 0.52 & 0.62 \\
    & Efficiency (\%) & 100.0 & 86.5 & 53.1 & 48.5 & 40.8 & 33.5 & 18.7 \\
    & Instances       & 1    & 1    & 1    & 1    & 1    & 1    & 1    \\
    & ($N_x, N_y$)    & (2,4)& (4,4)& (4,6)& (4,8)& (6,8)& (8,8)& (8,12)\\
    \midrule
    \multirow{4}{*}{\textbf{Scale-out}} 
    & Time (h)        & 1.36 & 0.81 & 1.02 & 0.85 & 0.86 & 0.69 & 0.82 \\
    & Efficiency (\%) & 100.0 & 83.4 & 44.3 & 39.8 & 26.3 & 24.6 & 13.8 \\
    & Instances       & 1    & 1    & 1    & 2    & 2    & 4    & 4    \\
    & ($N_x, N_y$)    & (2,4)& (4,4)& (4,6)& (4,8)& (6,8)& (8,8)& (8,12)\\
    \bottomrule
\end{tabularx}
\end{table}

\section{Conclusions}
\label{sec:conclusion}


Adviser enables domain scientists and other end users to run complex computational workloads on the cloud more easily and efficiently. 
With an intuitive UI/UX and the suite of features discussed in Section \ref{sec:design}, Adviser mitigates the critical barriers identified in Section \ref{sec:introduction} that have historically prevented widespread adoption of cloud computing and HPC. 
We demonstrated these capabilities through successful use of Adviser to run open-source computational glaciology software, Icepack and PISM. 
In future work, we plan to extend Adviser’s heuristics for price and performance optimization and increase durability for long-running jobs, further enhancing its utility for cloud-based HPC workflows.


\begin{acks}
D.H.\ was supported by Seeding Success and Scaling Success internal awards at Vanderbilt University.
S.C.\ and D.H.\ were supported by NSF Grant No.\ 2324735.
D.H., B.S., T.E., and K.W.\ disclose financial interest in Adviser Labs, Inc., which sells a commercial version of the Adviser software.
The authors wish to thank Tianrun Gao for assistance with proofreading and for preparing demos of Adviser. Perplexity Pro was used as a ``peer reviewer,'' and we used its feedback to improve the manuscript prior to submission. \texttt{llama3.3-70b-instruct} was used for the automated textual analysis discussed in Section \ref{sec:job-analysis}.
\end{acks}

\bibliographystyle{ACM-Reference-Format-etal3}
\bibliography{Adviser}

@article{chen2017running,
  title = {Running Climate Model on a Commercial Cloud Computing Environment: {{A}} Case Study Using {{Community Earth System Model}} ({{CESM}}) on {{Amazon AWS}}},
  author = {Chen, Xiuhong and Huang, Xianglei and Jiao, Chaoyi and Flanner, Mark G. and Raeker, Todd and Palen, Brock},
  year = 2017,
  month = jan,
  journal = {Computers \& Geosciences},
  volume = {98},
  pages = {21--25},
  issn = {00983004},
  doi = {10.1016/j.cageo.2016.09.014}
}

@article{madhavan2013nanohub,
  title = {{{nanoHUB}}.Org: Cloud-Based Services for Nanoscale Modeling, Simulation, and Education},
  author = {Madhavan, Krishna and Zentner, Lynn and Farnsworth, Victoria and Shivarajapura, Swaroop and Zentner, Michael and Denny, Nathan and Klimeck, Gerhard},
  year = 2013,
  month = feb,
  journal = {Nanotechnology Reviews},
  volume = {2},
  number = {1},
  pages = {107--117},
  issn = {2191-9097, 2191-9089},
  doi = {10.1515/ntrev-2012-0043}
}

@inproceedings{tran2022simvascular,
  title = {{{SimVascular Gateway}} for {{Education}} and {{Research}}},
  booktitle = {Practice and {{Experience}} in {{Advanced Research Computing}}},
  author = {Tran, Justin and Abeysinghe, Eroma and Ladisa, John and Marsden, Alison and Pierce, Marlon},
  year = 2022,
  month = jul,
  pages = {1--4},
  publisher = {ACM},
  address = {Boston MA USA},
  doi = {10.1145/3491418.3535162},
  isbn = {978-1-4503-9161-0}
}

@book{eynard2019pangeo,
  title = {{{Big Data}} from {{Space}}, {{Munich}} ({{Germany}}).},
  author = {{European Commission. Joint Research Centre.}},
  month = Feb,
  year = 2019,
  publisher = {Publications Office},
  address = {LU},
  doi = {10.2760/848593}
}

@article{zhuang2019enabling,
  title = {Enabling {{Immediate Access}} to {{Earth Science Models}} through {{Cloud Computing}}: {{Application}} to the {{GEOS-Chem Model}}},
  author = {Zhuang, Jiawei and Jacob, Daniel J. and Gaya, Judith Flo and Yantosca, Robert M. and Lundgren, Elizabeth W. and Sulprizio, Melissa P. and Eastham, Sebastian D.},
  year = 2019,
  month = oct,
  journal = {Bulletin of the American Meteorological Society},
  volume = {100},
  number = {10},
  pages = {1943--1960},
  issn = {0003-0007, 1520-0477},
  doi = {10.1175/BAMS-D-18-0243.1}
}

@article{zhuang2020enabling,
  title = {Enabling {{High}}-{{Performance Cloud Computing}} for {{Earth Science Modeling}} on {{Over}} a {{Thousand Cores}}: {{Application}} to the {{GEOS}}-{{Chem Atmospheric Chemistry Model}}},
  author = {Zhuang, Jiawei and Jacob, Daniel J. and Lin, Haipeng and Lundgren, Elizabeth W. and Yantosca, Robert M. and Gaya, Judit Flo and Sulprizio, Melissa P. and Eastham, Sebastian D.},
  year = 2020,
  month = may,
  journal = {Journal of Advances in Modeling Earth Systems},
  volume = {12},
  number = {5},
  pages = {e2020MS002064},
  issn = {1942-2466, 1942-2466},
  doi = {10.1029/2020MS002064}
}

@article{oreproducible,
  title = {Reproducible {{Workflow}} on a {{Public Cloud}} for {{Computational Fluid Dynamics}}},
  author = {Mesnard, Olivier and Barba, Lorena A.},
  year = 2020,
  month = jan,
  journal = {Computing in Science \& Engineering},
  volume = {22},
  number = {1},
  pages = {102--116},
  issn = {1521-9615, 1558-366X},
  doi = {10.1109/MCSE.2019.2941702},
  copyright = {https://ieeexplore.ieee.org/Xplorehelp/downloads/license-information/IEEE.html}
}

@inproceedings{bauer2012legion,
  title = {Legion: {{Expressing}} Locality and Independence with Logical Regions},
  booktitle = {2012 {{International Conference}} for {{High Performance Computing}}, {{Networking}}, {{Storage}} and {{Analysis}}},
  author = {Bauer, Michael and Treichler, Sean and Slaughter, Elliott and Aiken, Alex},
  year = 2012,
  month = nov,
  pages = {1--11},
  publisher = {IEEE},
  address = {Salt Lake City, UT},
  doi = {10.1109/SC.2012.71},
  isbn = {978-1-4673-0805-2 978-1-4673-0806-9}
}

@incollection{breuer2019petaflop,
  title = {Petaflop {{Seismic Simulations}} in the {{Public Cloud}}},
  booktitle = {High {{Performance Computing}}},
  author = {Breuer, Alexander and Cui, Yifeng and Heinecke, Alexander},
  editor = {Weiland, Mich{\`e}le and Juckeland, Guido and Trinitis, Carsten and Sadayappan, Ponnuswamy},
  year = 2019,
  volume = {11501},
  pages = {167--185},
  publisher = {Springer International Publishing},
  address = {Cham},
  doi = {10.1007/978-3-030-20656-7_9},
  isbn = {978-3-030-20655-0 978-3-030-20656-7}
}

@inproceedings {yang2023skypilot,
author = {Zongheng Yang and Zhanghao Wu and Michael Luo and Wei-Lin Chiang and Romil Bhardwaj and Woosuk Kwon and Siyuan Zhuang and Frank Sifei Luan and Gautam Mittal and Scott Shenker and Ion Stoica},
title = {{SkyPilot}: An Intercloud Broker for Sky Computing},
booktitle = {20th USENIX Symposium on Networked Systems Design and Implementation (NSDI 23)},
year = {2023},
isbn = {978-1-939133-33-5},
address = {Boston, MA},
pages = {437--455},
publisher = {USENIX Association},
month = apr
}

@article{shapero2021icepack,
  title = {Icepack: A New Glacier Flow Modeling Package in {{Python}}, Version 1.0},
  author = {Shapero, Daniel R. and Badgeley, Jessica A. and Hoffman, Andrew O. and Joughin, Ian R.},
  year = 2021,
  month = jul,
  journal = {Geoscientific Model Development},
  volume = {14},
  number = {7},
  pages = {4593--4616},
  issn = {1991-9603},
  doi = {10.5194/gmd-14-4593-2021},
  copyright = {https://creativecommons.org/licenses/by/4.0/}
}

@article{winkelmann2011potsdam,
  title = {The {{Potsdam Parallel Ice Sheet Model}} ({{PISM-PIK}}) -- {{Part}} 1: {{Model}} Description},
  author = {Winkelmann, R. and Martin, M. A. and Haseloff, M. and Albrecht, T. and Bueler, E. and Khroulev, C. and Levermann, A.},
  year = 2011,
  month = sep,
  journal = {The Cryosphere},
  volume = {5},
  number = {3},
  pages = {715--726},
  issn = {1994-0424},
  doi = {10.5194/tc-5-715-2011},
  copyright = {https://creativecommons.org/licenses/by/3.0/}
}

@article{martin2011potsdam,
  title = {The {{Potsdam Parallel Ice Sheet Model}} ({{PISM-PIK}}) -- {{Part}} 2: {{Dynamic}} Equilibrium Simulation of the {{Antarctic}} Ice Sheet},
  author = {Martin, M. A. and Winkelmann, R. and Haseloff, M. and Albrecht, T. and Bueler, E. and Khroulev, C. and Levermann, A.},
  year = 2011,
  month = sep,
  journal = {The Cryosphere},
  volume = {5},
  number = {3},
  pages = {727--740},
  issn = {1994-0424},
  doi = {10.5194/tc-5-727-2011},
  copyright = {https://creativecommons.org/licenses/by/3.0/}
}

@article{joughin2021ocean,
  title = {Ocean-Induced Melt Volume Directly Paces Ice Loss from {{Pine Island Glacier}}},
  author = {Joughin, Ian and Shapero, Daniel and Dutrieux, Pierre and Smith, Ben},
  year = 2021,
  month = oct,
  journal = {Science Advances},
  volume = {7},
  number = {43},
  pages = {eabi5738},
  issn = {2375-2548},
  doi = {10.1126/sciadv.abi5738}
}

@article{joughin2021ice,
  title = {Ice-Shelf Retreat Drives Recent {{Pine Island Glacier}} Speedup},
  author = {Joughin, Ian and Shapero, Daniel and Smith, Ben and Dutrieux, Pierre and Barham, Mark},
  year = 2021,
  month = jun,
  journal = {Science Advances},
  volume = {7},
  number = {24},
  pages = {eabg3080},
  issn = {2375-2548},
  doi = {10.1126/sciadv.abg3080},
  copyright = {https://creativecommons.org/licenses/by/4.0/}
}

@inproceedings{rothwell2022quantifying,
  title = {Quantifying the {{Impact}} of {{Advanced Web Platforms}} on {{High Performance Computing Usage}}},
  booktitle = {Practice and {{Experience}} in {{Advanced Research Computing}}},
  author = {Rothwell, Bradlee and Sgambati, Matthew and Evans, Garrick and Biggs, Brandon and Anderson, Matthew},
  year = 2022,
  month = jul,
  pages = {1--8},
  publisher = {ACM},
  address = {Boston MA USA},
  doi = {10.1145/3491418.3530758},
  isbn = {978-1-4503-9161-0}
}

@incollection{ahrens2005paraview,
  title = {{{ParaView}}: {{An End-User Tool}} for {{Large-Data Visualization}}},
  booktitle = {Visualization {{Handbook}}},
  author = {Ahrens, James and Geveci, Berk and Law, Charles},
  year = 2005,
  pages = {717--731},
  publisher = {Elsevier},
  address = {Amsterdam, The Netherlands},
  doi = {10.1016/B978-012387582-2/50038-1},
  copyright = {https://www.elsevier.com/tdm/userlicense/1.0/},
  isbn = {978-0-12-387582-2}
}

@misc{saturn-cloud,
  title   = {Saturn Cloud: Code-First {AI} Infrastructure},
  author  = {{Saturn Cloud}},
  year    = {2026},
  month   = jan,
  url     = {https://saturncloud.io/},
  urldate = {2026-01-19}
}

@misc{nvidia-run-ai,
  author  = {{NVIDIA}},
  title   = {NVIDIA {Run:ai}: Accelerate {AI} \& Machine Learning Workflows},
  year    = {2026},
  month   = jan,
  url     = {https://www.nvidia.com/en-us/software/run-ai/},
  urldate = {2026-01-19}
}

@misc{modal-stack,
  author  = {{Modal}},
  title   = {Modal: High-performance {AI} Infrastructure},
  year    = {2026},
  month   = jan,
  url     = {https://modal.com/},
  urldate = {2026-01-19}
}

@inproceedings{upcxx_ipdps19,
  title = {{{UPC}}++: {{A High-Performance Communication Framework}} for {{Asynchronous Computation}}},
  booktitle = {{{International Parallel}} and {{Distributed Processing Symposium}}},
  author = {Bachan, John and Baden, Scott B. and Hofmeyr, Steven and Jacquelin, Mathias and Kamil, Amir and Bonachea, Dan and Hargrove, Paul H. and Ahmed, Hadia},
  year = 2019,
  month = may,
  pages = {963--973},
  publisher = {IEEE},
  address = {Rio de Janeiro, Brazil},
  doi = {10.1109/IPDPS.2019.00104},
  copyright = {https://ieeexplore.ieee.org/Xplorehelp/downloads/license-information/USG.html},
  isbn = {978-1-7281-1246-6}
}

@inproceedings{ompss2cluster_europar22,
  author    = {Aguilar Mena, Jimmy and Shaaban, Omar and Beltran, Vicen{\c{c}} and Carpenter, Paul and Ayguade, Eduard and Labarta Mancho, Jesus},
  title     = {{OmpSs-2}@{Cluster}: Distributed Memory Execution of Nested OpenMP-Style Tasks},
  booktitle = {Euro-Par 2022: Parallel Processing},
  year      = {2022},
  pages     = {319--334},
  publisher = {Springer International Publishing},
  address   = {Cham},
  doi       = {10.1007/978-3-031-12597-3_20},
  url       = {https://doi.org/10.1007/978-3-031-12597-3_20}
}

@inproceedings{nosv_ipdps24,
  title = {{{nOS-V}}: {{Co-Executing HPC Applications Using System-Wide Task Scheduling}}},
  booktitle = {2024 {{IEEE International Parallel}} and {{Distributed Processing Symposium}} ({{IPDPS}})},
  author = {{\'A}lvarez, David and Sala, Kevin and Beltran, Vicen{\c c}},
  year = 2024,
  month = may,
  pages = {312--324},
  publisher = {IEEE},
  address = {San Francisco, CA, USA},
  doi = {10.1109/IPDPS57955.2024.00035},
  copyright = {https://doi.org/10.15223/policy-029},
  isbn = {979-8-3503-8711-7}
}

@article{argobots_tpds17,
  title = {Argobots: {{A Lightweight Low-Level Threading}} and {{Tasking Framework}}},
  author = {Seo, Sangmin and Amer, Abdelhalim and Balaji, Pavan and Bordage, Cyril and Bosilca, George and Brooks, Alex and Carns, Philip and Castello, Adrian and Genet, Damien and Herault, Thomas and Iwasaki, Shintaro and Jindal, Prateek and Kale, Laxmikant V. and Krishnamoorthy, Sriram and Lifflander, Jonathan and Lu, Huiwei and Meneses, Esteban and Snir, Marc and Sun, Yanhua and Taura, Kenjiro and Beckman, Pete},
  year = 2018,
  month = mar,
  journal = {IEEE Transactions on Parallel and Distributed Systems},
  volume = {29},
  number = {3},
  pages = {512--526},
  issn = {1045-9219},
  doi = {10.1109/TPDS.2017.2766062},
  copyright = {https://ieeexplore.ieee.org/Xplorehelp/downloads/license-information/IEEE.html}
}

@incollection{irisreimagined_wamta24,
  title = {{{IRIS Reimagined}}: {{Advancements}} in {{Intelligent Runtime System}} for {{Task-Based Programming}}},
  booktitle = {Asynchronous {{Many-Task Systems}} and {{Applications}}},
  author = {Miniskar, Narasinga Rao and Lee, Seyong and Beau, Johnston and Young, Aaron and Monil, Mohammad Alaul Haque and {Valero-Lara}, Pedro and Vetter, Jeffrey S.},
  editor = {Diehl, Patrick and Schuchart, Joseph and {Valero-Lara}, Pedro and Bosilca, George},
  year = 2024,
  volume = {14626},
  pages = {46--58},
  publisher = {Springer Nature Switzerland},
  address = {Cham},
  doi = {10.1007/978-3-031-61763-8_5},
  isbn = {978-3-031-61762-1 978-3-031-61763-8}
}

@article{numrich1998co,
  title = {Co-Array {{Fortran}} for Parallel Programming},
  author = {Numrich, Robert W. and Reid, John},
  year = 1998,
  month = aug,
  journal = {ACM SIGPLAN Fortran Forum},
  volume = {17},
  number = {2},
  pages = {1--31},
  issn = {1061-7264, 1931-1311},
  doi = {10.1145/289918.289920}
}

@techreport{carlson1999introduction,
  title={Introduction to UPC and language specification},
  author={Carlson, William W and Draper, Jesse M and Culler, David E and Yelick, Kathy and Brooks, Eugene and Warren, Karen},
  year={1999},
  institution={Technical Report CCS-TR-99-157, IDA Center for Computing Sciences}
}

@inproceedings{stubbs2021tapis,
  title = {Tapis: {{An API}} Platform for Reproducible, Distributed Computational Research},
  booktitle = {Advances in Information and Communication},
  author = {Stubbs, Joe and Cardone, Richard and Packard, Mike and Jamthe, Anagha and Padhy, Smruti and Terry, Steve and Looney, Julia and Meiring, Joseph and Black, Steve and Dahan, Maytal and Cleveland, Sean and Jacobs, Gwen},
  editor = {Arai, Kohei},
  year = 2021,
  pages = {878--900},
  publisher = {Springer International Publishing},
  address = {Cham},
  isbn = {978-3-030-73100-7}
}

@article{swetnam2024cyverse,
  title = {{{CyVerse}}: {{Cyberinfrastructure}} for Open Science},
  author = {Swetnam, Tyson L. and Antin, Parker B. and Bartelme, Ryan and Bucksch, Alexander and Camhy, David and Chism, Greg and Choi, Illyoung and Cooksey, Amanda M. and Cosi, Michele and Cowen, Cindy and {Culshaw-Maurer}, Michael and Davey, Robert and Davey, Sean and Devisetty, Upendra and Edgin, Tony and Edmonds, Andy and Fedorov, Dmitry and Frady, Jeremy and Fonner, John and Gillan, Jeffrey K. and Hossain, Iqbal and Joyce, Blake and Lang, Konrad and Lee, Tina and Littin, Shelley and McEwen, Ian and Merchant, Nirav and Micklos, David and Nelson, Andrew and Ramsey, Ashley and Roberts, Sarah and Sarando, Paul and Skidmore, Edwin and Song, Jawon and Sprinkle, Mary Margaret and Srinivasan, Sriram and Stanzione, Dan and Strootman, Jonathan D. and Stryeck, Sarah and Tuteja, Reetu and Vaughn, Matthew and Wali, Mojib and Wall, Mariah and Walls, Ramona and Wang, Liya and Wickizer, Todd and Williams, Jason and Wregglesworth, John and Lyons, Eric},
  editor = {Papin, Jason A.},
  year = 2024,
  month = feb,
  journal = {PLOS Computational Biology},
  volume = {20},
  number = {2},
  pages = {e1011270},
  issn = {1553-7358},
  doi = {10.1371/journal.pcbi.1011270}
}

@inproceedings{lentner2019shared,
author = {Lentner, Geoffrey},
title = {Shared Memory High Throughput Computing with Apache Arrow™},
year = {2019},
isbn = {9781450372275},
publisher = {Association for Computing Machinery},
address = {New York, NY, USA},
url = {https://doi.org/10.1145/3332186.3335197},
doi = {10.1145/3332186.3335197},
booktitle = {Practice and Experience in Advanced Research Computing 2019: Rise of the Machines (Learning)},
articleno = {119},
numpages = {2},
location = {Chicago, IL, USA},
series = {PEARC '19}
}

@article{rathgeber2016firedrake,
author = {Rathgeber, Florian and Ham, David A. and Mitchell, Lawrence and Lange, Michael and Luporini, Fabio and Mcrae, Andrew T. T. and Bercea, Gheorghe-Teodor and Markall, Graham R. and Kelly, Paul H. J.},
title = {Firedrake: Automating the Finite Element Method by Composing Abstractions},
year = {2016},
issue_date = {September 2017},
publisher = {Association for Computing Machinery},
address = {New York, NY, USA},
volume = {43},
number = {3},
issn = {0098-3500},
url = {https://doi.org/10.1145/2998441},
doi = {10.1145/2998441},
journal = {ACM Trans. Math. Softw.},
month = dec,
articleno = {24},
numpages = {27}
}

@article{deelman2015pegasus,
  title = {Pegasus, a Workflow Management System for Science Automation},
  author = {Deelman, Ewa and Vahi, Karan and Juve, Gideon and Rynge, Mats and Callaghan, Scott and Maechling, Philip J. and Mayani, Rajiv and Chen, Weiwei and Ferreira Da Silva, Rafael and Livny, Miron and Wenger, Kent},
  year = 2015,
  month = may,
  journal = {Future Generation Computer Systems},
  volume = {46},
  pages = {17--35},
  issn = {0167739X},
  doi = {10.1016/j.future.2014.10.008}
}

@article{di2017nextflow,
  title = {Nextflow Enables Reproducible Computational Workflows},
  author = {Di Tommaso, Paolo and Chatzou, Maria and Floden, Evan W and Barja, Pablo Prieto and Palumbo, Emilio and Notredame, Cedric},
  year = 2017,
  month = apr,
  journal = {Nature Biotechnology},
  volume = {35},
  number = {4},
  pages = {316--319},
  issn = {1546-1696},
  doi = {10.1038/nbt.3820}
}

@article{koster2012snakemake,
  title = {Snakemake---a Scalable Bioinformatics Workflow Engine},
  author = {K{\"o}ster, Johannes and Rahmann, Sven},
  year = 2012,
  month = oct,
  journal = {Bioinformatics},
  volume = {28},
  number = {19},
  pages = {2520--2522},
  issn = {1367-4811, 1367-4803},
  doi = {10.1093/bioinformatics/bts480}
}

@inproceedings{babuji2019parsl,
author = {Babuji, Yadu and Woodard, Anna and Li, Zhuozhao and Katz, Daniel S. and Clifford, Ben and Kumar, Rohan and Lacinski, Lukasz and Chard, Ryan and Wozniak, Justin M. and Foster, Ian and Wilde, Michael and Chard, Kyle},
title = {Parsl: Pervasive Parallel Programming in Python},
year = {2019},
isbn = {9781450366700},
publisher = {Association for Computing Machinery},
address = {New York, NY, USA},
url = {https://doi.org/10.1145/3307681.3325400},
doi = {10.1145/3307681.3325400},
booktitle = {Proceedings of the 28th International Symposium on High-Performance Parallel and Distributed Computing},
pages = {25–36},
numpages = {12},
location = {Phoenix, AZ, USA},
series = {HPDC '19}
}



\end{document}